\documentclass[prl,aps,showkeys,showpacs,superscriptaddress,twocolumn,amsmath,amssymb]
{revtex4-1} 
\usepackage{graphics}
\usepackage{bm}
\usepackage{subfigure}
\usepackage{epsfig}
\usepackage{epstopdf}
\usepackage[bookmarks,colorlinks]{hyperref}
\usepackage{natbib}
\usepackage{siunitx}
\newcommand{\beginsupplement}{%
        \setcounter{table}{0}
        \renewcommand{\thetable}{S\arabic{table}}%
        \setcounter{figure}{0}
        \renewcommand{\thefigure}{S\arabic{figure}}%
     }
\usepackage{xr}
\usepackage{placeins}
\begin{document}
\title{From yielding to shear jamming in a cohesive frictional suspension }
\author{Abhinendra Singh}
\email{asingh.iitkgp@gmail.com}
\affiliation{Benjamin Levich Institute, CUNY City College of New York, New York, NY 10031, USA}
\author{Sidhant Pednekar}
\email{sidhant11@gmail.com}
\affiliation{Benjamin Levich Institute, CUNY City College of New York, New York, NY 10031, USA}
\affiliation{Department of Chemical Engineering, CUNY City College of New York, New York, NY 10031, USA}
\affiliation{Pacific Northwest National Laboratory Richland, Washington 99352, USA}
\author{Jaehun Chun}
\email{jaehun.chun@pnnl.gov}
\affiliation{Pacific Northwest National Laboratory Richland, Washington 99352, USA}
%
\author{Morton M. Denn}
\email{denn@ccny.cuny.edu}
\affiliation{Benjamin Levich Institute, CUNY City College of New York, New York, NY 10031, USA}
\affiliation{Department of Chemical Engineering, CUNY City College of New York, New York, NY 10031, USA}
\author{Jeffrey F. Morris}
\email{morris@ccny.cuny.edu}
\affiliation{Benjamin Levich Institute, CUNY City College of New York, New York, NY 10031, USA}
\affiliation{Department of Chemical Engineering, CUNY City College of New York, New York, NY 10031, USA}
\date{\today} 
\begin{abstract}
Simulations are used to study the steady shear rheology of dense suspensions of frictional particles exhibiting discontinuous shear thickening and shear jamming, in which 
finite--range cohesive interactions result in a yield stress.
We develop a constitutive model that combines yielding behavior and shear thinning at low stress with the frictional shear thickening at high stresses, in good agreement with the simulation results.
%
%
%
This work shows that there is a distinct difference between solids below the yield stress and in the shear-jammed state, as the two occur at widely separated stress levels, separated by a region of stress in which the material is flowable.
\end{abstract}
\pacs{83.80.Hj, 83.60.Rs, 83.10.Rs}
\maketitle

{\it Introduction}:
Concentrated or ``dense'' suspensions of particles in liquid are found in both natural \cite{Coussot_1997} and industrial settings~\cite{Bridgwater_1993, Chun_2011, Peterson_2018}.
Under shear, non-Brownian suspensions display a number of non-Newtonian properties; considering just the shear properties, these mixtures may undergo yielding, shear thinning, shear thickening or even jamming~\cite{mewis_colloidal_2011, guazzelli_physical_2011, Denn_2014, Denn_2018}.
Such non-Newtonian rheology arises from particle interactions~\cite{Denn_2014}, influenced by the solid-fluid interfacial chemistry and chemical physics of both phases~\cite{Brown_2010,Brown_2014,Galvez_2017}, as well as from frictional interactions between particles~\cite{Comtet_2017,Fernandez_2013,Hsu_2018} that are influenced by roughness \cite{Lootens_2005,Hsiao_2017}.
Suspensions of particles interacting by attractive forces can exhibit a yield stress and at larger stresses shear thicken, and, as discussed here, possibly jam. 
Shear thickening (ST), the increase of relative viscosity $\eta_{\rm r}$ with increasing shear rate $\dot{\gamma}$, can occur as continuous shear thickening (CST) or discontinuous shear thickening (DST) in dense suspensions; here the relative viscosity is normalized by the suspending fluid viscosity $\eta_0$, $\eta_{\rm r} = \eta(\phi,\dot{\gamma})/\eta_0$, where $\phi$ is the volume fraction. 
The viscosity varies continuously with $\dot{\gamma}$ in CST, while DST is characterized by $d\eta_{\rm r}/d\dot{\gamma}\rightarrow \infty$ at some stress, often resulting in orders of magnitude increase in viscosity.
 It has been demonstrated that if $\phi$ is sufficiently large, the suspension can even become a shear-jammed (SJ) solid~\cite{Peters_2016}; this solid is fragile, in the sense that it is maintained in this state by the imposed load, and would, for example, fail if the load is applied in the reverse direction~\cite{Cates_1998a}.
A recent body of work~\cite{Wyart_2014,Mari_2014,Seto_2013a,Lin_2015,Guy_2015,mari_discontinuous_2015} has related shear thickening to a transition from lubricated to frictional interactions of particles above an ``onset stress.'' 
An approach capturing this two-state model~\cite{Wyart_2014} based on a mean-field description of the fraction of particle interactions that are frictional has been shown~\cite{Singh_2018} to be successful in describing both the relative viscosity $\eta_{\rm r}$ and normal stress differences found in simulations of shear thickening frictional suspensions.

To date, most study has been focused on the flow behavior of dense non-cohesive suspensions. However, van der Waals forces~\cite{Maranzano_2001a}, depletion forces due to dissolved non-interacting polymer~\cite{Gopalakrishnan_2004}, or the presence of an external field~\cite{Brown_2010} can all lead to attractive forces between particles. 
A demonstrated influence of attractive forces is that the shear thickening may be obscured~\cite{Gopalakrishnan_2004,Brown_2010,Pednekar_2017}.  When the low-stress viscosity becomes sufficiently large or a yield stress develops, a suspension shear thins to a high shear-rate viscosity, which in the case of the shear-thickening suspension would be the thickened state of the non-cohesive suspension~\cite{Gopalakrishnan_2004,Brown_2010,Pednekar_2017,Galvez_2017}. 
\par
The studies noted just above addressed volume fractions exhibiting CST. 
It is our particular goal to demonstrate the influence of cohesion for suspensions of volume fractions for which the non-cohesive suspension undergoes DST and SJ. 
The latter will illustrate that the same material may exhibit shear yielding at low stress, flow at intermediate stress, and shear jam at high stress.
This  provides a distinctly different picture of the relation of yielding and jamming than has been suggested in other work \cite{LiuNagel_AnnRev}, as these two phenomena occur at widely separated stress levels, separated by a region of stresses for which the material is flowable. 

We explore a broad range of volume fractions, with a focus on $\phi$ close to the frictional jamming volume fraction, denoted $\phi_{\rm J}^\mu$.
In this range of solid fraction, non-cohesive suspensions show DST and shear jamming. We extend a constitutive model for dense frictional suspension rheology~\cite{Singh_2018} to cohesive systems exhibiting yielding and shear-thinning in addition to shear thickening.
Using the simulation results and guided by this model, a state diagram for dense frictional suspensions with attractive interactions is proposed.

{\it Simulations:}
 An assembly of inertialess spheres suspended in an equal density Newtonian fluid is simulated, under conditions of imposed shear stress $\sigma$, as described previously \cite{Mari_2015}. The suspension flows at a time-dependent shear rate $\dot\gamma(t)$ in a 3D Lees-Edwards periodic computational domain. We simulate 500 particles in the domain, using equal volume fractions of particle radii $a$ and $1.4a$. The bidispersity avoids ordering.  Simulations with 2000 particles have been performed to test finite-size effects.
 \par
The particles interact through short-range hydrodynamic lubrication forces $F_H$, a conservative force $F_{\mathrm {cons}} = F_A+F_R$ (where $A$ and $R$ denote the attractive and repulsive contributions, respectively), and contact forces $F_C$.  The contact force allows friction, with friction coefficient $\mu$. 
An electrostatic repulsion force decaying with interparticle surface separation $h$ over a length scale defined by Debye length $\lambda$ is used: 
$|F_R|  = F_0 \exp(-h/\lambda)$. To model the force of attraction, a van der Waals form  $F_A(h) = A \bar{a}/12(h^2+H^2)$ is used, where $A$ denotes the Hamaker constant and $\bar{a}$ denotes the harmonic mean radius $\bar{a} = 2a_1a_2/(a_1+a_2)$~\cite{Russel_1992}. The parameter $H$ is fixed at $H=0.1\bar{a}$, and is employed to eliminate the divergence of $F_A$ at contact ($h=0$). The conservative force is illustrated in Fig.\ref{force_profile}. The strength of attraction is controlled by $A$, which determines the value of the attractive force at contact, $F_A(0)$ (referred to as $F_A$ in the rest of the article).
The contact force between two particles is modeled by linear springs and dashpots as described elsewhere \cite{Mari_2014}. Tangential and normal components of the contact force $F_C$ between two particles satisfy the Coulomb friction law  $|F_{C,t}| \le \mu|F_{C,n}|$, where $\mu=1$ is used in the current work (note that $F_{C,n}$ is only compressive here.)

{\it Simulation results:}
\begin{figure}
\centering
\subfigure{
\includegraphics[width=.25\textwidth]{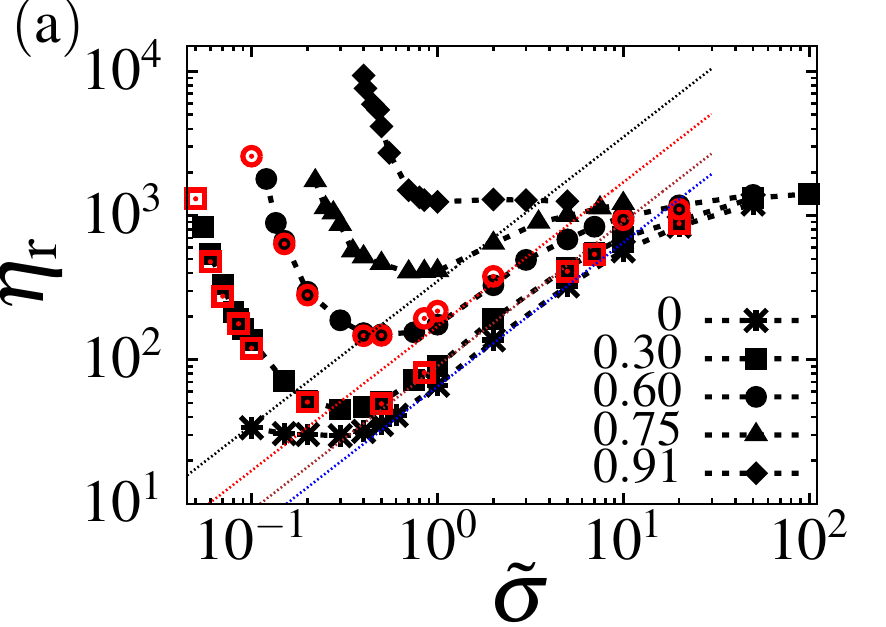}}\hspace*{-1.2em}
\subfigure{
\includegraphics[width=.25\textwidth]{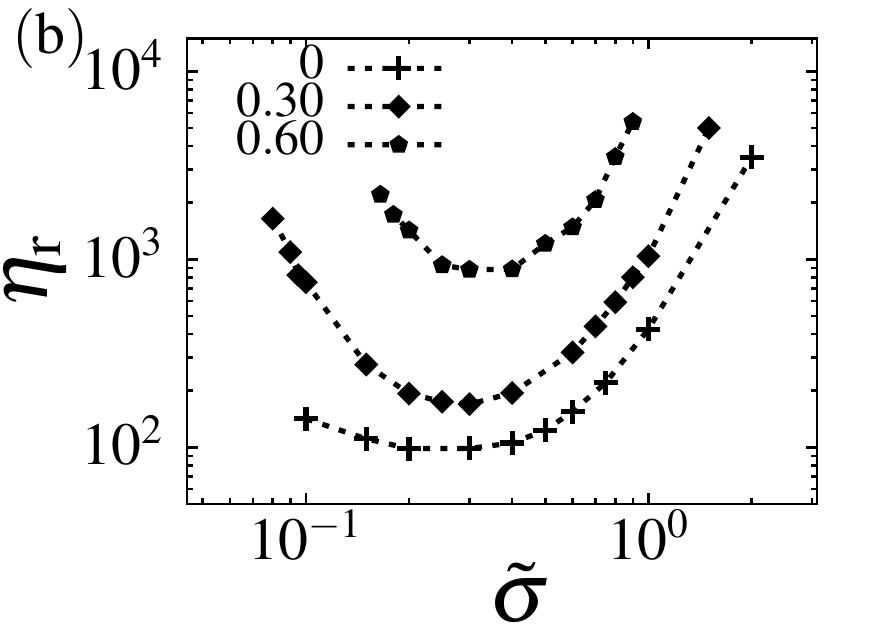}}
\caption{%
Relative viscosity $\eta_{\rm r}$ plotted versus dimensionless applied stress $\tilde{\sigma} = {\sigma}/\sigma_0$ for volume fraction $\phi=$ (a) 0.56 and (b) 0.6 and several values of attractive strength $F_A$ at $\mu=1$.
The symbols are simulation data with dashed lines provided to guide the eye.
%
%
Open (red) symbols in (a) are results with (N=2000) for $F_A=$ 0.3 and 0.6, showing results are very similar and finite size effects are minimal for the conditions studied here.
Dotted lines in (a) show $\eta_{\rm r} \propto {\sigma}/\sigma_0$, signifying increase in viscosity at constant rate, as shown in Fig.~\ref{fig:visc_rate}a.
%
%
}
\label{fig:visc_fa0p56}
\end{figure}
Figure~\ref{fig:visc_fa0p56} shows the influence of attractive forces on the rheology of a frictional non-Brownian suspension for $\phi = 0.56$ and 0.6, where the shear stress is scaled by $\sigma_0=F_0/6\pi a^2$ (using the smaller particle radius).
To characterize the steepness of the viscosity increase in the $\eta_r$ {\em vs.} $\sigma/\sigma_0$ flow curve, the shear--thickening portion is fitted to $\eta_{\rm r} \propto ({\sigma}/\sigma_0)^\beta$, where $\beta<1$ signifies CST and $\beta=1$ indicates that the shear rate, $\dot{\gamma}/\dot{\gamma}_0 = \eta_{\rm r}/({\sigma}/\sigma_0)$, is unchanging while stress increases and hence is the onset of DST.
 For $\phi=0.56$, the non-cohesive frictional suspension shows DST between two flowing states, as is evident from $\eta_{\rm r} \propto {\sigma}/\sigma_0$ (i.e. $\beta = 1$) in Fig.~\ref{fig:visc_fa0p56}a.
%
The development of a moderate yield stress $\sigma_{\rm y}$ was observed for $F_A=0.3$. 
For $F_A\ge 0.3$, the suspension flows when $\sigma>\sigma_{\rm y}$, first shear thinning from the infinite viscosity of the unyielded material and 
eventually shear thickening.  This thickening begins continuously, but as $\sigma$ is further increased DST occurs. 
An increase in $F_A$ increases $\sigma_y$, which by raising the minimum viscosity reached by shear thinning 
effectively weakens the extent of shear thickening. For $F_A=0$ to 0.6, discontinuous shear thickening is still observed, as shown by dotted lines showing $\eta_{\rm r} \propto {\sigma}/\sigma_0$. 
Development of a yield stress, indicated by a  slope of -1 in Fig.~\ref{fig:visc_rate}a,  does not immediately lead to obscuring of shear thickening. However for  $F_A=0.75$, DST is not observed, as only a weak shear thickening is needed to carry the suspension from its minimum viscosity to the high-stress plateau. All shear thickening is obscured with further increase in $F_A$, consistent with previous simulation and experimental studies at lower volume fractions~\cite{Pednekar_2017, Gopalakrishnan_2004, Brown_2010}. 
 %
 
At $\phi=0.6$, exceeding the frictional jamming fraction, $\phi_{\rm J}^{\mu}\approx 0.585$ for $\mu = 1$ as shown elsewhere \cite{Singh_2018}, the suspension shear jams at sufficiently large shear stress, $\sigma_{\rm sj}(\phi)$. Upon introducing cohesion, the suspension develops a yield stress $\sigma_{\rm y}$ and cannot flow for $\sigma< \sigma_{\rm y}$. Thus, the cohesive frictional suspension is a non--flowable solid for $\sigma<\sigma_{\rm y}$, flows at intermediate stress, and shear jams above $\sigma_{\rm {sj}}$. However, for $F_A=0.91$ the suspension cannot flow for any value of shear stress, as $\sigma_{\rm y}> \sigma_{\rm sj}$.  Note that below the yield stress one has a standard, albeit soft, solid that resists deformation in all directions equally if prepared without directional bias, whereas the shear-jammed solid at $\sigma>\sigma_{\rm sj}$ is fragile in the absence of the cohesive forces, and thus has anisotropic properties~\cite{Cates_1998a, Bi_2011}. 
%
%
 
{\it Constitutive model:}
\begin{figure}
\centering
\subfigure{
\includegraphics[width=.25\textwidth]{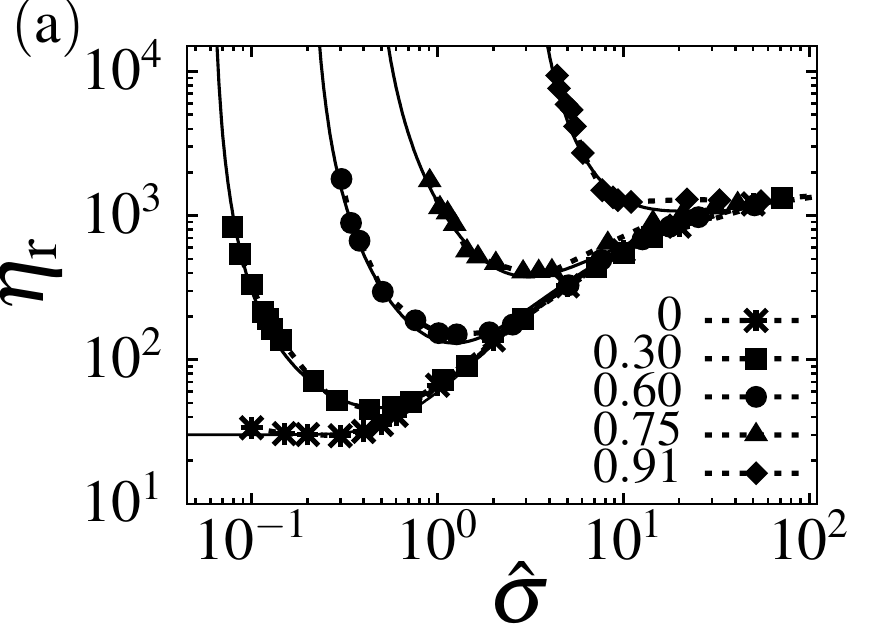}}\hspace*{-1.2em}
\subfigure{
\includegraphics[width=.25\textwidth]{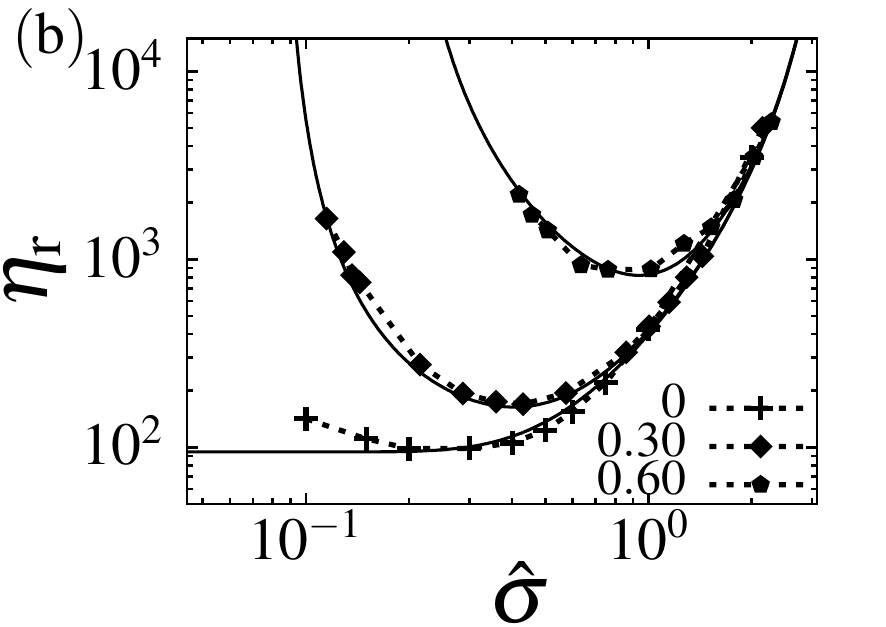}}
\caption{Steady state flow curves for volume fractions $\phi=0.56$ and 0.6 from Figs.~\ref{fig:visc_fa0p56}a and~\ref{fig:visc_fa0p56}c plotted versus scaled applied shear stress $\hat{\sigma}  = \sigma/\sigma_0^{\rm {RA}}$ where
$\sigma_0^{\rm {RA}} =\left[ F_A(0)+F_R(0) \right] /{6\pi a^2}$; where $F_R(0)$ and $F_A(0)$ are the values of repulsive and attractive forces, respectively, at surface separation $h=0$. The symbols are simulations with different values of strength of
attraction and the solid lines are predictions from~\eqref{eq:eta_total}.
}
\label{fig:visc_fa}
\end{figure}
We renormalize the scaled stress as $\hat{\sigma} = \sigma/\sigma_0^{\rm {RA}}$ and shear rate as $\hat{\dot{\gamma}} = \hat{\sigma}/\eta_0$, where the scaling factor $\sigma_0^{\rm {RA}} = (F_A+F_R)/{6\pi a^2}$ is the sum of repulsive and attractive stress magnitudes at surface separation $h=0$. 
%
In Fig.~\ref{fig:visc_fa} we plot steady state viscosity $\eta_{\rm r}$ versus shear stress $\hat{\sigma} $. 
We observe the collapse of viscosity data for intermediate to high stress to the non-cohesive flow curve. The yield stress decreases the range of stresses for which shear thickening is observed.

To quantify the effect of attractive interactions on the flow behavior of shear thickening suspensions, we use the Herschel-Bulkley 
equation 
\begin{equation}
\hat{\sigma}_{\mathrm {HB}}(\hat{\dot{\gamma}}) = \hat{\sigma}_{\rm y}+K\hat{\dot{\gamma}}^{n},
\label{eq:HB}
\end{equation}
where $\hat{\sigma}_{\rm y}$ denotes the scaled yield stress, $K$ is the consistency index and $n$ is the power law exponent. We find that $n=0.5$ describes the yielding and shear--thinning behavior well for all $F_A$ and $\phi$ considered here, consistent with prior  studies~\cite{Maranzano_2001,Zaccone_2011,Brown_2010,Dagois_2015}. We recast Eq.~\eqref{eq:HB} as 
\begin{equation}\label{eq:eta_HB}
\eta_{\rm r}^{\rm {HB}} (\hat{\sigma})= \frac{K^2\hat{\sigma}_{\rm y}}{(\hat{\sigma} -\hat{\sigma}_{\rm y})^2} + \frac{K^2 }{(\hat{\sigma} - \hat{\sigma}_{\rm y})}~.
\end{equation}
The model parameters $\hat{\sigma}_{\rm y}$ and $K$ are obtained by fitting the low stress (yielding and shear-thinning) portion of the flow curve to Eq.~\eqref{eq:eta_HB}.
 The shear thickening of non-cohesive suspension viscosity has been expressed as~\cite{Singh_2018} 
 \begin{subequations}\label{eq:eta_phi_str}
 \begin{equation}
\eta_{\rm r}^{\rm C} (\phi,\hat{\sigma}) = \alpha_{\rm m}(\hat{\sigma}) [\phi_{\rm m} (\hat{\sigma}) - \phi]^{-2} ~,
\end{equation}
where 
 \begin{equation}
 \phi_{\rm m}(\hat{\sigma}) = \phi_{\rm J}^\mu  f(\hat{\sigma}) + \phi_{\rm J}^0 [ 1-f(\hat{\sigma}) ]~
 \end{equation}
 and 
 \begin{equation}
 \alpha_{\rm m}(\hat{\sigma}) =  \alpha^\mu f(\hat{\sigma}) + {\alpha^0} (1-f(\hat{\sigma}))
 \end{equation}
 \end{subequations}
 interpolate between two values of $\phi$ and $\alpha$, while $f \in [0, 1]$ represents the fraction of frictional contacts, whose form is presented in Mari {\em et al.} \cite{Mari_2014}.
As in earlier works~\cite{Brown_2010,Zaccone_2011, Gopalakrishnan_2004}, various contributions to the viscosity can be superimposed as
\begin{equation}\label{eq:eta_total}
\eta_{\rm r}(\phi,\hat{\sigma}) = \eta_{\rm r}^{\rm {HB}} (\phi,\hat{\sigma}) + \eta_{\rm r}^{\rm C} (\phi,\hat{\sigma}).
\end{equation}
The viscosity modeled by Eq.~\eqref{eq:eta_total} is compared to the simulation data in Fig.~\ref{fig:visc_fa} and is seen to agree well. 
We also find that the second normal stress difference $N_2$ (shown in Supplementary Material, Fig.~\ref{fig-N2coh}) behaves in a fashion similar to the shear stress, i.e., it displays cohesion-dependent yield behavior at low stress while the behavior is independent of attraction at high stress.
In Fig.~\ref{fig-brownnoncoh} we demonstrate a possible extension of the model to non--cohesive Brownian suspensions capturing well both Brownian shear-thinning and frictional shear-thickening.

{\it Origin of yielding:}
\begin{figure}
\centering
\subfigure{
\includegraphics[width=.25\textwidth]{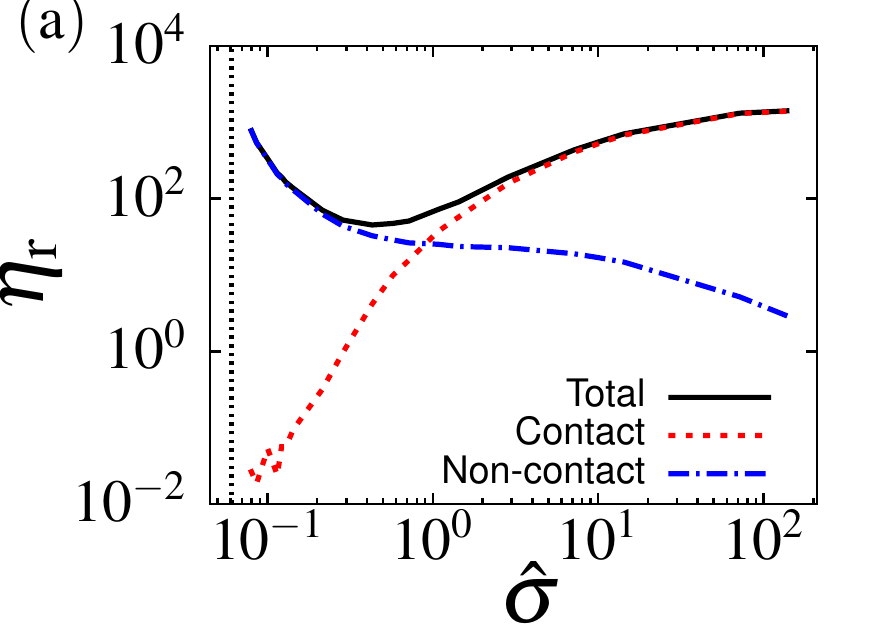}}\hspace*{-1.2em}
\subfigure{
\includegraphics[width=.25\textwidth]{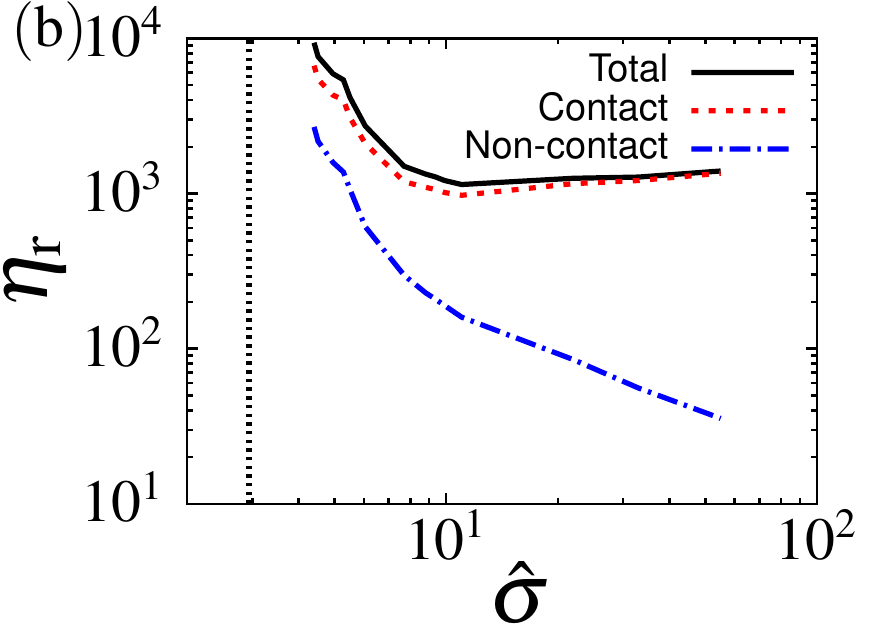}}
\caption{Total relative viscosity $(\eta_{\rm r})$ and the contributions arising from
hydrodynamic interactions, conservative forces, and contact forces, plotted
versus scaled shear stress $\hat{\sigma}$ for a non-Brownian suspension $(\phi=0.56)$ with $F_A=$ (a) 0.3 and (b) 0.91.
The broken vertical line indicates the yield stress.
%
}
\label{fig-2}
\end{figure}
In an attempt to get a more mechanistic understanding of the behavior of cohesive shear-thickening systems, we focus on the origin of yielding, especially with increasing force of attraction. Subsequently, we separate the total viscosity into contact and non-contact contributions which are shown as functions of stress $\hat{\sigma}$ for $F_A=0.3$ and 0.91 in Figs.~\ref{fig-2}a and~\ref{fig-2}b, respectively. The hydrodynamic contribution to overall viscosity is insignificant for the conditions presented. At low strengths of attraction, non-contact (attractive and repulsive) forces provide the dominant contribution to overall viscosity at low stresses while the contact contribution takes over at higher stresses. Snapshots of force networks for $F_A=0.3$ at low stress are plotted in Fig.~\ref{fig-network}a. Particles are seen to interact only via finite-range (non-contact) forces. On closer inspection, repulsive forces are seen to interact primarily along the compressive axis as they resist approaching particles. Attractive forces, by contrast, generate resistance along the extensional axis for departing particles.

In stark contrast, high strengths of cohesion result in a dominant contribution from contact forces, irrespective of stress (as seen in Fig.~\ref{fig-2}b). Figure~\ref{fig-network}b confirms this through the presence of frictional force networks in the system even at low stress. Figure \ref{fig:visc_fa}a provides insight into this behavior; the yield stress for $F_A=0.91$ is larger than the onset stress $\hat{\sigma}_{\rm {on}}\doteq 0.3$ for the non--cohesive ($F_A=0$) curve. Stronger attractive forces dominate over repulsive forces in this regime, bringing particles into contact to allow formation of the frictional force networks seen in Fig.~\ref{fig-network}b. Frictional contacts are capable of resisting applied shear stress, leading to an increase in yield stress and viscosity. 

{\it Flow state diagram}:
%
\begin{figure}
\centering
\includegraphics[width=.45\textwidth]{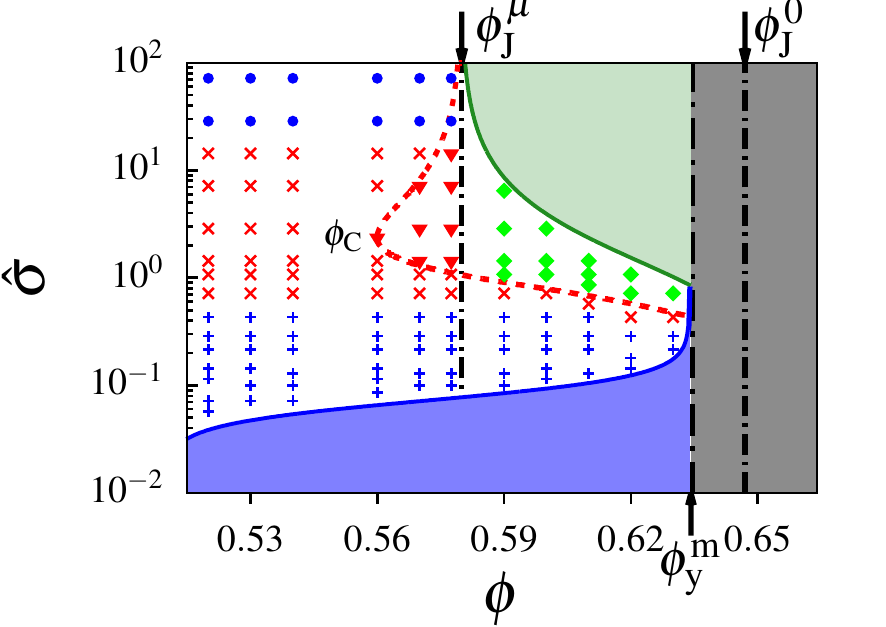}
%
\caption{ 
Flow state diagram in $\hat{\sigma} - \phi$ plane for $F_A=0.3$ showing shear jammed (green), unyielded (blue), flowing (white) and inaccessible (gray) states. The green and blue solid lines are the stress--dependent jamming and yield lines, respectively, while the red dashed line is the DST line and shows the locus of points where $d\dot{\gamma}/d\sigma=0$. Dot--dashed black lines show $\phi_{\rm J}^\mu$, $\phi_{\rm y}^{\rm m}$ and $\phi_{\rm J}^0$. Symbols show different flowing states of the suspension: Shear thinning (blue pluses),
 shear thickened state (blue circles), CST (red crosses), DST between two flowing states (red inverted triangles), DST between a flowing and a jammed state (green diamonds). Here, we have only probed volume fractions $\phi \ge 0.52$, and the yield line might continue for lower volume fractions. 
%
}
\label{fig:phase_diagram_fa}
\end{figure}
%
Using the simulation results of the present work (data in Fig.~\ref{fig-eta_FA}), we construct a flow state diagram in the $\hat{\sigma} -\phi$ plane, as shown in Figure~\ref{fig:phase_diagram_fa} for $F_A=0.3$. Since the focus of the present study is on the rheological behavior for volume fraction close to DST or above, we have only probed volume fractions $\phi \ge 0.52$.
For the range of volume fraction $\phi_{\rm J}^\mu<\phi<\phi_{\rm y}^{\rm m}$ the suspension is in different solid states for $\hat{\sigma}<\hat{\sigma}_{\rm y}$ and $\hat{\sigma}>\hat{\sigma}_{\rm {sj}}$.
There is a volume fraction $\phi_{\rm y}^{\rm m}$ above which flow does not occur at any stress $\hat{\sigma}$. 
With increasing $\phi$ the range of stress $\hat{\sigma}$ for which the system can flow shrinks until it vanishes at $\phi_{\rm y}^{\rm m}$.
%
For $\phi < \phi_{\rm J}^\mu$ the system is in an unyielded solid state for $\sigma<\sigma_{\rm y}$ with flow at larger stress. 
 The yield stress increases with $\phi$ and diverges at $\phi_{\rm y}^{\rm m}$, which is smaller than $\phi_{\rm J}^0$; this behavior has also been observed previously for other non-Brownian suspensions~\cite{Zhou_1995}.
For volume fractions below $\phi_{\rm C}$, continuous shear thickening (CST) is observed for intermediate stress values.
For $\phi_{\rm C} \le \phi < \phi_{\rm J}^{\mu}$, DST is observed between two flowing states as shown by the dashed (red) line which is the locus of points where $d\dot{\gamma}/d\sigma=0$,
while for $\phi_{\rm J}^{\mu} < \phi < \phi_{\rm y}^{\rm m}$, the upper boundary of DST states is the stress-dependent jamming line $\phi_{\rm m}(\hat{\sigma})$ shown by the solid (green) line. A similar flow-state diagram was proposed recently~\cite{Guy_2018} using constraint counting arguments.
%
Since we present the state diagram for a single nonzero $F_A$, we note that at $F_A=0$, the state diagram would reduce to the one proposed previously~\cite{Singh_2018}. Higher values of $F_A$ could result in $\phi_{\rm y}^{\rm m} < \phi_{\rm J}^\mu$,
  thereby obscuring shear jamming.

Equations~\ref{eq:eta_HB} and~\ref{eq:eta_total} demonstrate how the development of yield stress and shear thinning shrinks the range of stress for which shear thickening is observed. 
For a given volume fraction $\phi$, increasing $F_A$ leads to an increase in yield stress $\hat{\sigma}_{\rm y}$, which in turn increases both $\hat{\sigma}_{\mathrm {on}}$ and the viscosity at the onset of shear thickening. 
The  viscosity at $\hat{\sigma}_{\mathrm {on}}$ should follow $\eta_{\rm r}(\hat{\sigma}_{\mathrm {on}},\phi) \le \eta_{\rm r}^\mu(\phi)$, where $\eta_{\rm r}^\mu(\phi)$ is the viscosity of the thickened (frictional) state.
At the equality shear thickening is obscured, implying that the system yields directly to  the frictional branch.

{\it Conclusions:} In this work we have studied the rheology of dense suspensions interacting through both finite-range cohesive and frictional contact interactions. We report flow curves that show yielding behavior at low stress and shear thickening as well as jamming at high stress, depending on the volume fraction $\phi$ relative to its frictional jamming value $\phi_{\rm J}^\mu$. This yield-to-jamming within a single 
concentration suspension has been conceptualized ~\cite{Wagner_2009}, but never previously reported from experiment or simulation. 
This behavior provides a clear distinction between yielding and jamming, unlike other  suggestions of these being essentially the same ~\cite{LiuNagel_AnnRev}.
We have proposed a constitutive model that captures the observed behavior.
The yield stress $\hat{\sigma}_{\rm y}$ depends on the strength of attraction, which in principle be controlled by particle size, microstructure, chemistry at solid-fluid interfaces, and properties of fluid and solid phases such as dielectric properties~\cite{Scales_1998, Maranzano_2001, Zhou_2001, Brown_2010, Galvez_2017}.

Our work thus provides fundamental insight into the complex rheological behavior of particle suspensions based on balances between shearing, conservative and frictional forces.
Although we have used specific force profiles for the repulsive and attractive forces, namely electrostatic repulsion and van der Waals cohesion, the modeling of rheology of dense suspensions, and the proposed state diagram, should be qualitatively similar for generic attractive and repulsive forces.
Additionally, the proposed state diagram can, in principle, be extended to encompass systems shear thinning of Brownian viscosity.
%
%
%

{\it Acknowledgments:}   
We thank Ryohei Seto for useful discussions. 
%
This work was supported, in part, under National Science Foundation Grants CNS-0958379, CNS-0855217, ACI-1126113 and the City University of New York High Performance Computing Center at the College of Staten Island. JFM was supported by NSF 1605283.
Discussion on particle forces (JC) was supported by the Interfacial Dynamics in Radioactive Environments and Materials (IDREAM), an Energy Frontier Research Center funded by the U.S. Department of Energy (DOE), Office of Science, Basic Energy Sciences.

\bibliography{dst} 
\bibliographystyle{apsrev4-1}

\clearpage

\begin{widetext}

\begin{appendix}


\section*{\large Supplemental Material for ``From yielding to shear jamming in a cohesive frictional suspension"}
In this document we provide details about the conservative forces used in the simulations. Plots on the origin of yielding for intermediate cohesion strength are also provided towards a greater comprehension of the yielding and shear thickening behavior. The parameters used in the modelling of the effect of cohesion on shear thickening are presented, along with a possible extension of the model to Brownian suspensions. Also discussed is the dependence of second normal stress difference on strength of cohesion.
\beginsupplement
\section*{Force profile}
Figure~\ref{force_profile} shows the force profile of conservative forces used in this study. We only show the lowest and highest strengths of attractive forces. Dotted and dot--dashed lines show the attractive $(F_A(h) = A\bar{a}/12(h^2+H^2))$ and repulsive ($|F_R|  = F_0 \exp(-h/\lambda)$) force components, respectively. Solid black and red (gray) lines show the resulting total force against interparticle surface-surface separations for the lowest $(F_{\rm A}=0.3)$ and highest $(F_{\rm A}=0.91)$ strengths of attraction.
\begin{figure}[ht]
\includegraphics[width=0.45\linewidth]{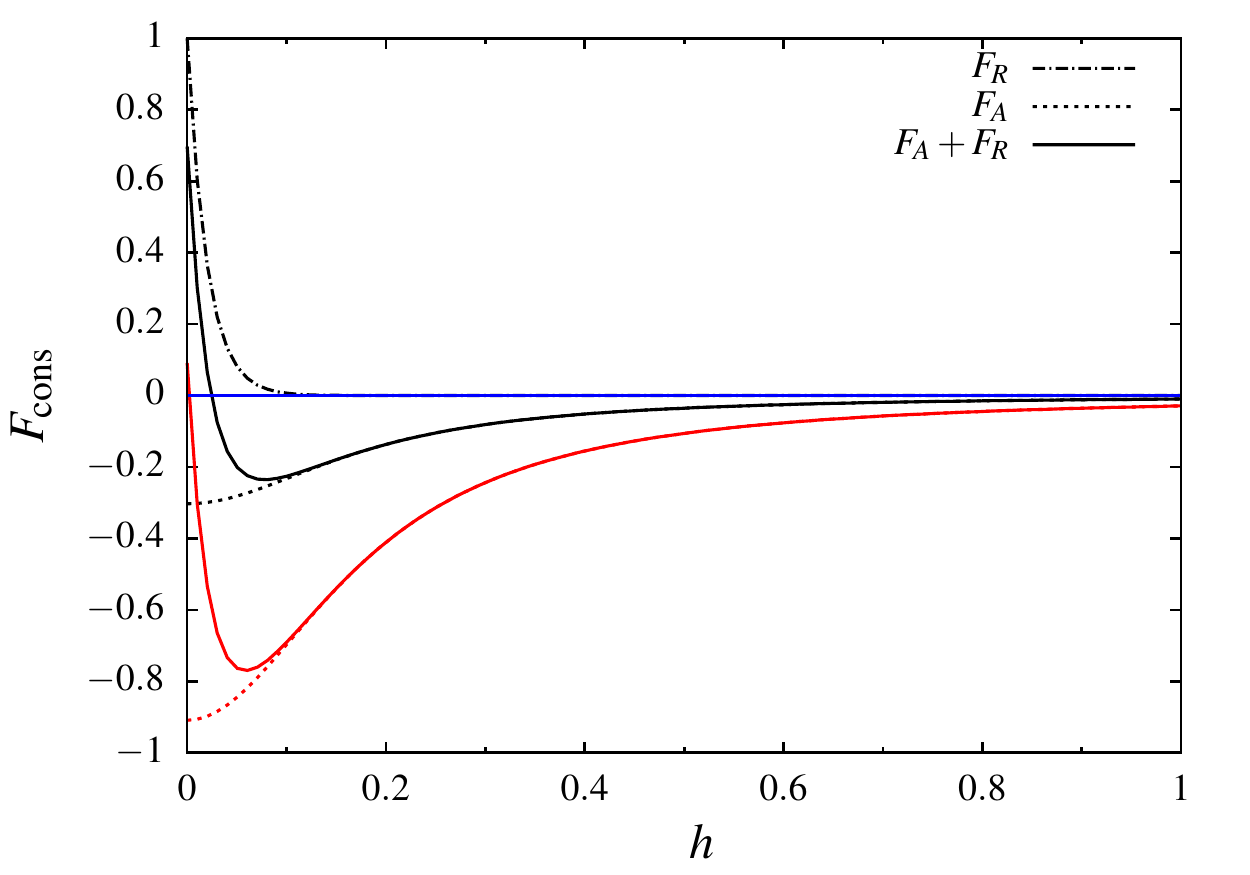}
\caption{
Conservative force function plotted as a function of the surface separation. Dotted line shows van der Waals attractive force $F_A(h) = A\bar{a}/12(h^2+H^2)$, while dot-dashed line shows repulsive force $|F_R|  = F_0 \exp(-h/\lambda)$.
Solid line displays the superposition of the two forces. Black and red colored lines show the lowest $(F_{\rm A}=0.3)$ and highest $(F_{\rm A}=0.91)$ strengths of attraction studied here, while blue lines is a reference to zero.
} 
\label{force_profile}
\end{figure}

\clearpage

\section*{Detailed rheological analysis for $\phi=0.56$}
Figure~\ref{fig:visc_rate} shows the steady state relative viscosity plotted as a function of shear rate, for the same data presented in Fig.~\ref{fig:visc_fa0p56}.
Here shear rate is scaled by $\sigma_0/\eta_0$, with $\sigma_0$ as defined in the main text.
For $\phi=0.56$, the viscosity is found to vary with shear stress as $\dot{\gamma}/\dot{\gamma}_0 = \eta_{\rm r}/({\sigma}/\sigma_0)$ signifying an increase in viscosity for unchanging shear rate
as shown in Fig.~\ref{fig:visc_rate}a. 
The flow curve is sigmoidal for $\phi=0.6>\phi_{\rm J}^{\mu}$ with $\dot{\gamma} \rightarrow 0$ signifying jamming.

\begin{figure}[!ht]
\centering
\subfigure{
\includegraphics[width=.5\textwidth]{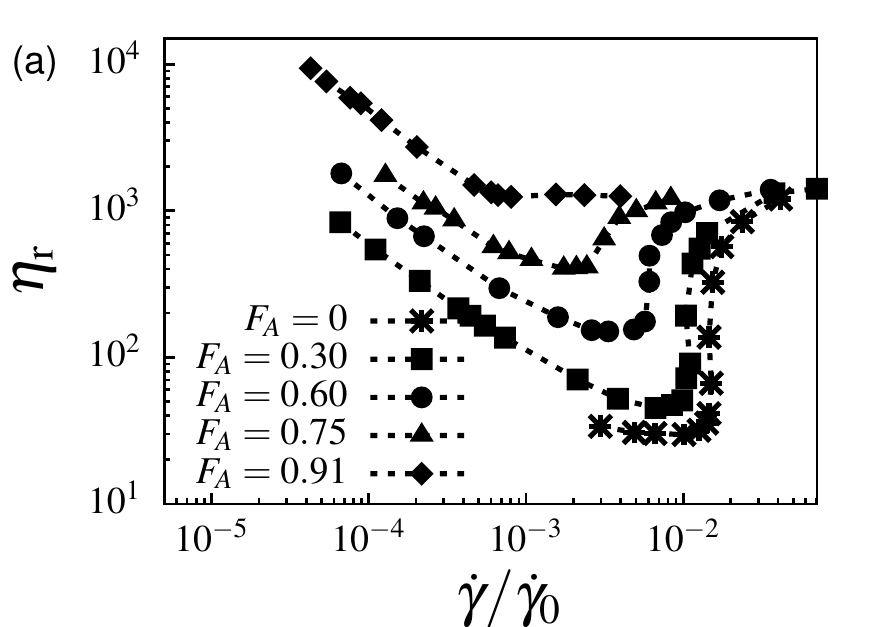}}
\subfigure{
\includegraphics[width=.5\textwidth]{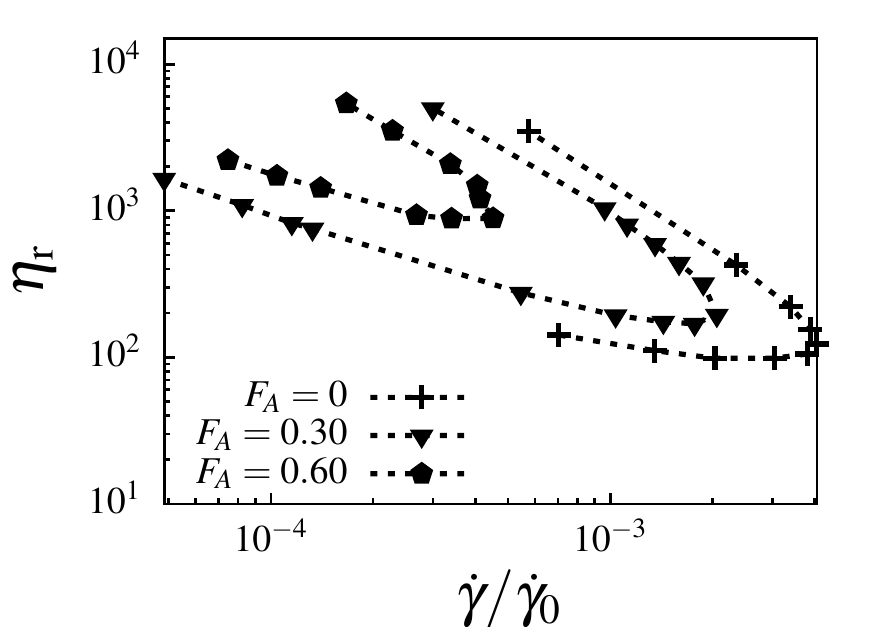}}
\caption{%
Relative viscosity $\eta_{\rm r}$ as a function of dimensionless shear rate $\dot{\gamma}/\dot{\gamma}_0$ for volume fraction $\phi=$ (a) 0.56 and (b) 0.6 and several values of attractive strength $F_A$ at $\mu=1$.
The symbols are simulation data with dashed lines provided to guide the eye.
The slope in a logarithmic plot of the viscosity as a function of shear rate approaching $-1$ implies yield stress.
}
\label{fig:visc_rate}
\end{figure}

Figure~\ref{flowcurve_errorbar} shows the steady state relative viscosity plotted as a function of stress with error bars (standard deviation in the data) for two different system sizes.
For $N=500$ (filled symbols), the error bars are large close to the yield stress $\sigma_{\rm y}$. With an increase in system size to $N=2000$, the mean value of viscosity
is similar, but fluctuations are reduced due to better statistics.

\begin{figure}[!ht]
\includegraphics[width=0.45\linewidth]{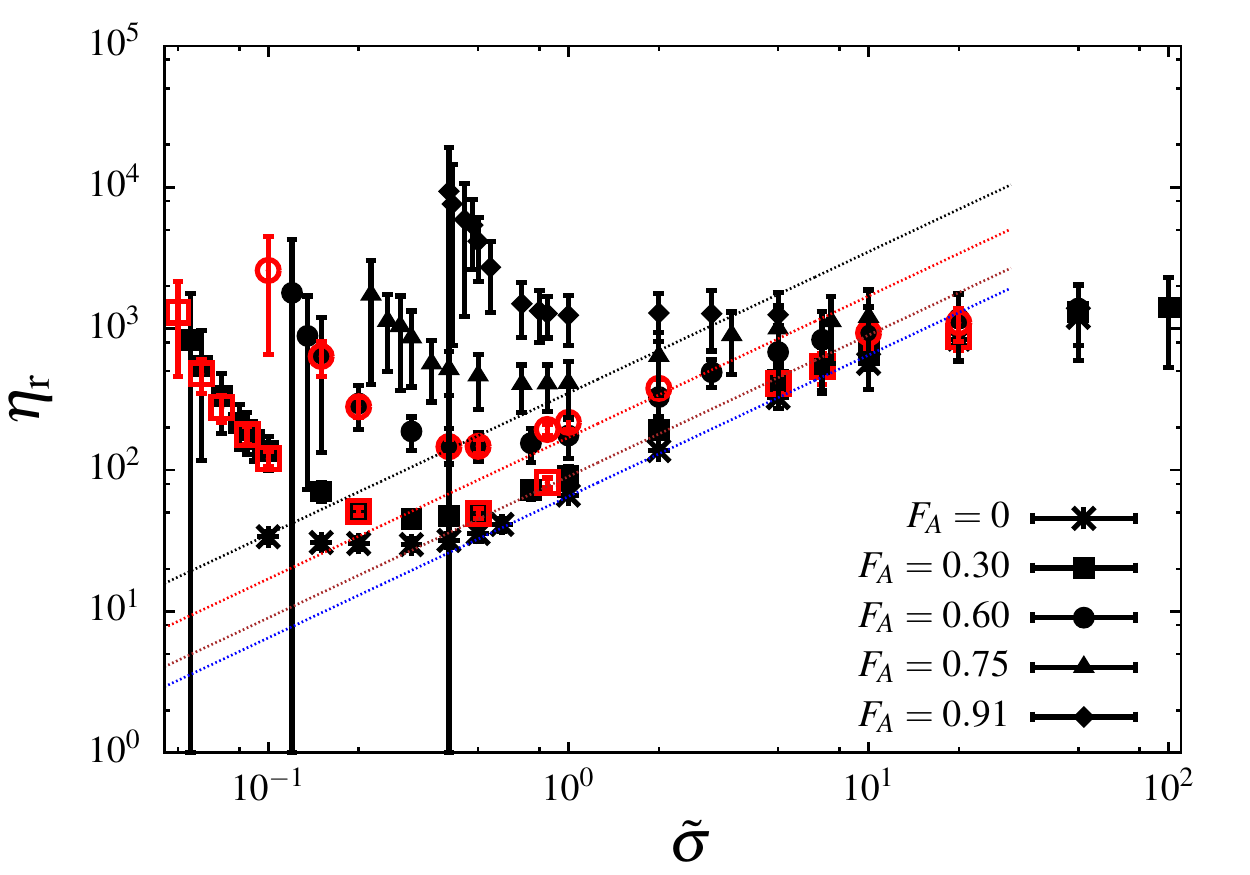}
\caption{
Detailed steady state flow curves with error bars for volume fraction $\phi=0.56$ for the data presented in Fig.~\ref{fig:visc_fa}a.
Error bars represent the standard deviation in the data.
Open (red) symbols in are results with N=2000 showing mean values are very similar and increasing the system size drastically reduces the variance in the data.
} 
\label{flowcurve_errorbar}
\end{figure}

\clearpage

\section{Force network for $\phi=0.56$ for $F_A=0.3$ and 0.91}
 Figure~\ref{fig-network} shows force network snapshots for the onset of flow for $\phi=0.56$ for the lowest and highest strengths of cohesion $F_A$ studied.
 It shows the dominance of non--contact interactions (attractive and repulsive) at low strength of attraction, but sufficient to result in a yield stress.
 On the other hand for $F_A=0.91$ frictional contacts develop and resist applied shear leading to the origin of yield stress.
\begin{figure}[!ht]
\centering
\subfigure[]{
\includegraphics[width=.45\textwidth]{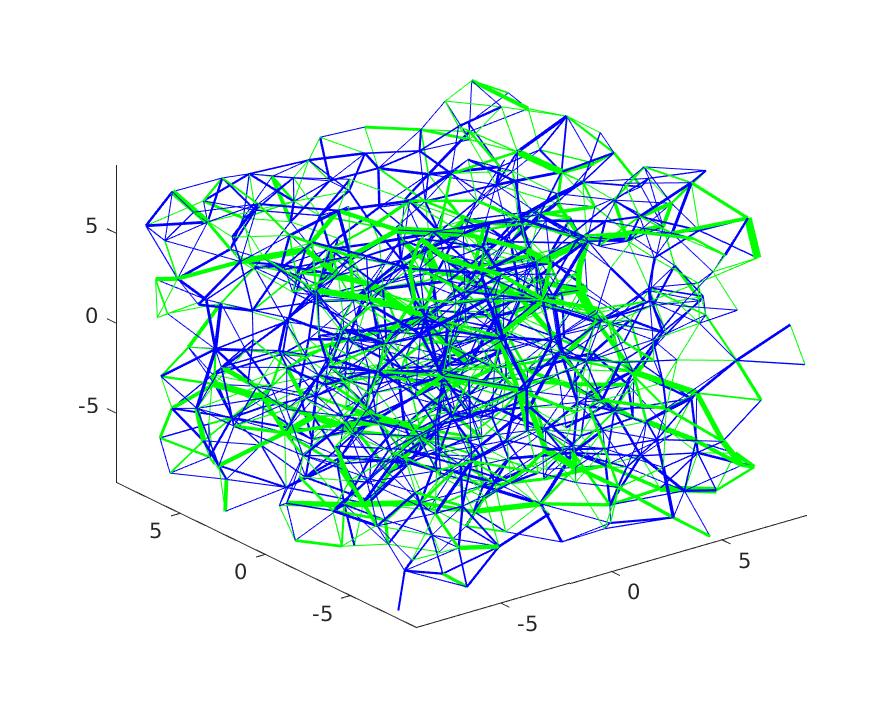}}
\subfigure[]{
\includegraphics[width=.45\textwidth]{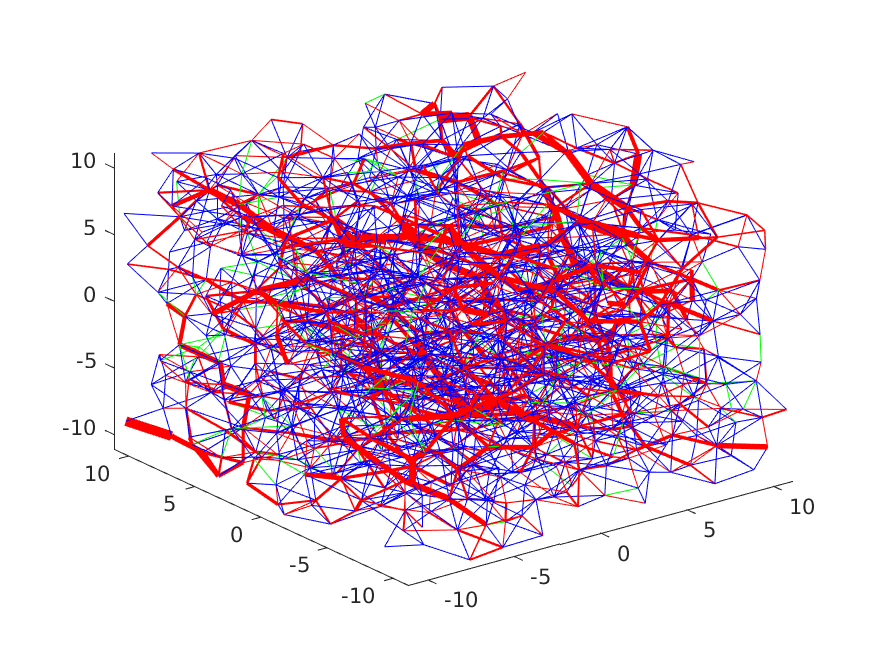}}
\caption{
Force network snapshots at onset of flow at $\phi=0.56$ for different
attractive strengths $F_A=$ (a) 0.3 and (b) 0.91. The thickness of the lines
are proportional to the normalized force ($\tilde{f}=f/\tilde{\sigma}a^2$) with red lines signifying direct frictional contacts;
green lines showing repulsive interactions; 
while blue lines show attractive interactions.
}
\label{fig-network}
\end{figure}

\clearpage

\section*{Origin of yielding for intermediate strength of cohesion for $\phi=0.56$}
 
Figure~\ref{fig-origin} displays the contact and non--contact contributions to viscosity along with the total viscosity as a function of scaled stress. 
The data plotted here along with the results for lowest and highest strengths of cohesion (shown in Fig.~\ref{fig-2}) clearly
 show the increase in the contact contribution to the viscosity. 
 With increasing strength of attraction, the origin of yielding crosses over from an attraction that is non--contact dominated to frictional contact domination.
\begin{figure*}[h]
\centering
 \subfigure[]{
 \includegraphics[width=.45\textwidth]{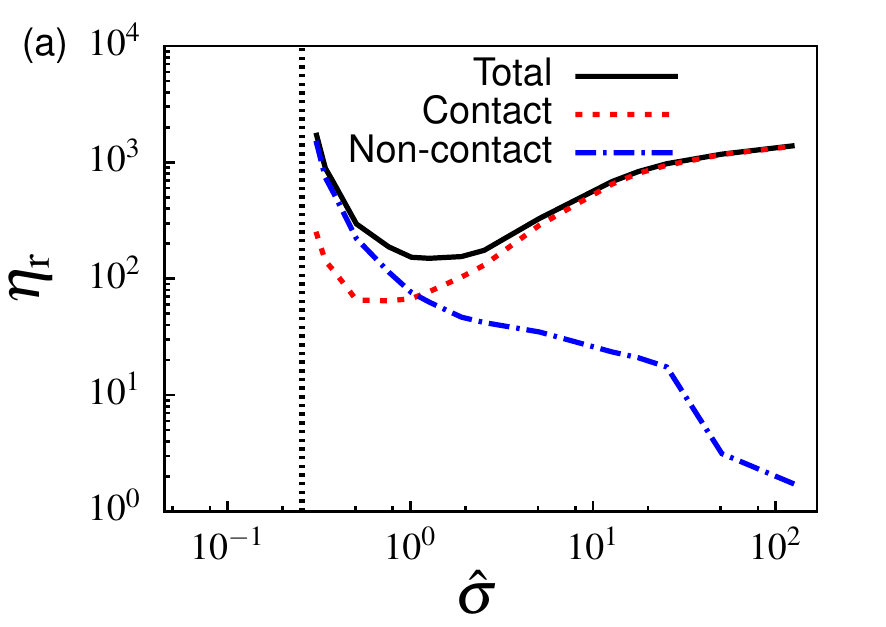}}
 \subfigure[]{
 \includegraphics[width=.45\textwidth]{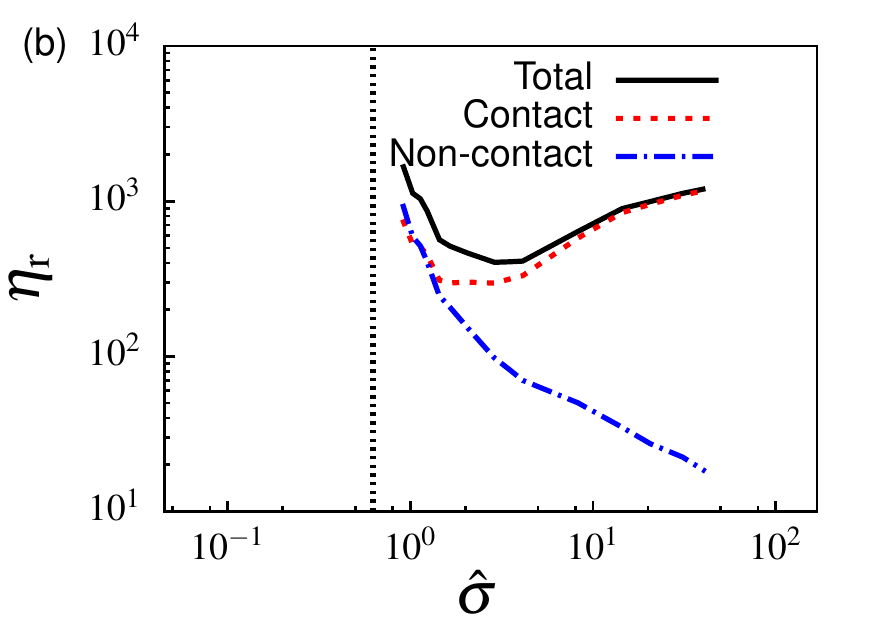}}
\caption{
Total relative viscosity $(\eta_r)$ and the contributions arising from
hydrodynamic interactions, conservative forces, and contact forces, plotted
as a function of scaled shear stress $\hat{\sigma}$ for non-Brownian suspension $(\phi=0.56)$ with $F_A=$ (a) 0.6 and (b) 0.76.
Dotted vertical line shows the yield stress.
%
}
\label{fig-origin}
\end{figure*}
\clearpage

\section*{Yield stress $\hat{\sigma}_{\rm y}$ and consistency index $K$}

Figure~\ref{fig-consistency} displays $\hat{\sigma}_{\rm y}$ and consistency  $K$ plotted against volume fraction $\phi$ for simulation data for $F_A=0.3$
showing that both $\hat{\sigma}_{\rm y}$ and $K$ have a weak dependence on $\phi$ for low volume fractions $(\phi \le 0.54)$, consistent with previous findings~\cite{Brown_2010};
the dependence is seen to increase at higher $\phi$ and diverge at $\phi_{\rm y}^{\rm m}$.
\begin{figure}[!ht]
\centering
\subfigure[]{
\includegraphics[width=.45\textwidth]{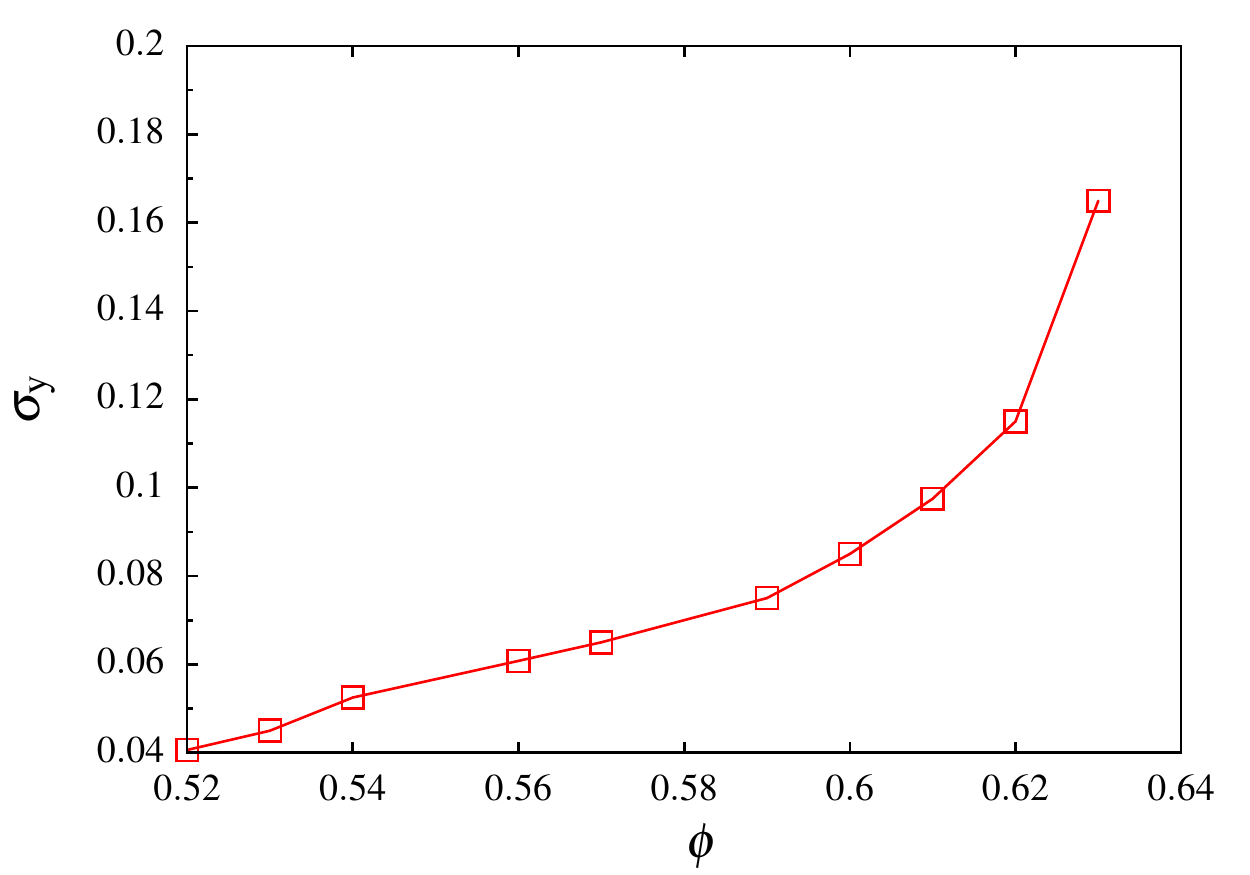}}
\subfigure[]{
\includegraphics[width=.45\textwidth]{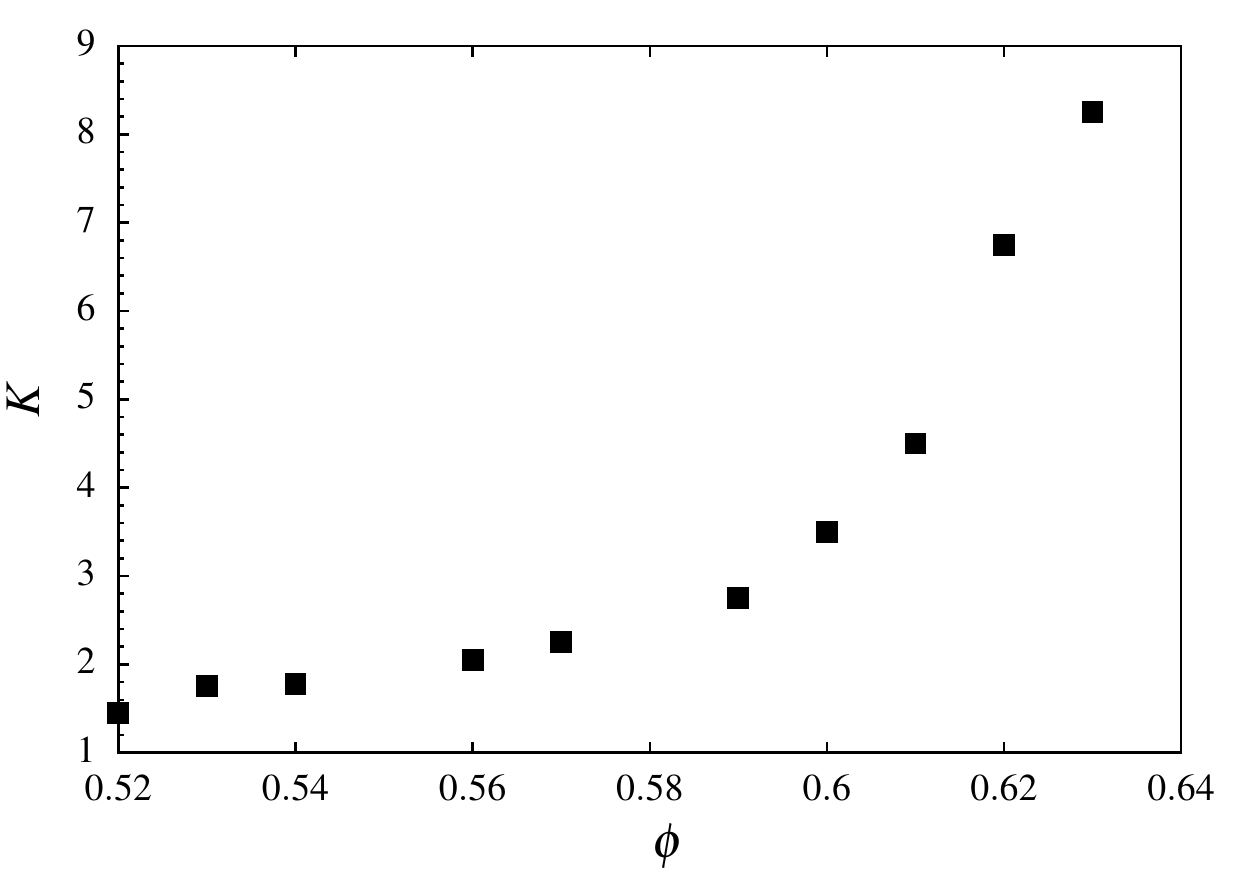}}
\caption{ Fit parameters (a) yield stress $\hat{\sigma}_{\rm y}$ and (b) consistency $K$ obtained by fitting low stress data to Herschel--Bulkley equation plotted against volume fraction $\phi$ for $F_A=0.3$ and $\mu=1$.
%
}
\label{fig-consistency}
\end{figure}


\clearpage

\section*{Extension of model for non--cohesive Brownian suspension}

Non-cohesive Brownian suspensions undergo shear thinning at low shear. Using a model as presented in Eq. \eqref{eq:HB}, which captures shear thinning from a fitted yield stress, we propose a possible extension to Brownian systems. In Figure~\ref{fig-brownnoncoh} we employ the Brownian simulation data from Mari {\it {et al.}}~\cite{mari_discontinuous_2015}. We plot relative viscosity $\eta_{\rm r}$ against stress rescaled by the onset stress $\sigma_{\rm on}=5k_BT/a^3+0.01F_0/a^2$ fordifferent repulsion amplitudes $F_0$. The solid line shows the prediction of the model, which agrees well with the simulation results of Mari {\it {et al.}}~\cite{mari_discontinuous_2015}.

\begin{figure*}[h]
\centering
 \includegraphics[width=.45\textwidth]{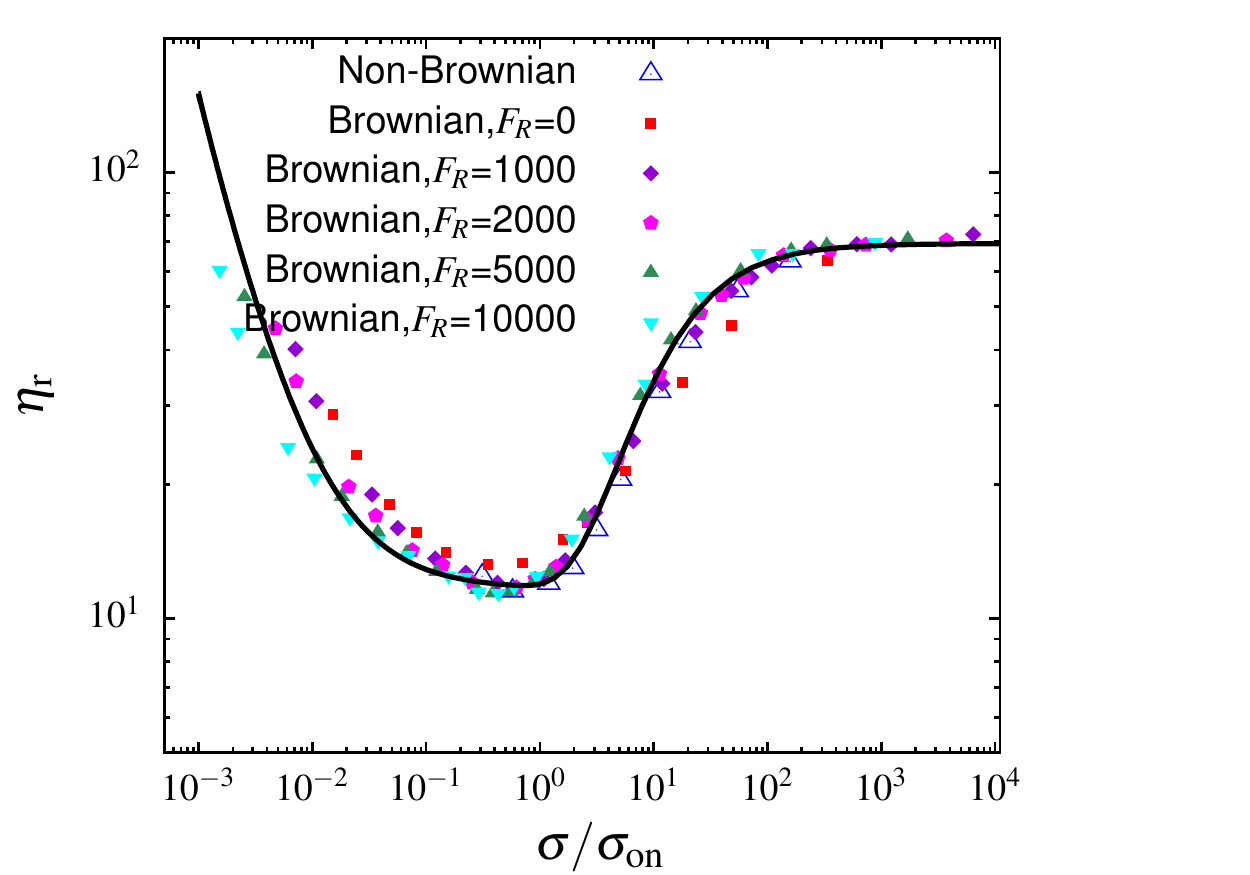}
\caption{
Symbols: relative viscosity vs rescaled stress $\sigma/\sigma_{\rm {on}}$ for non--cohesive Brownian suspension data from Ref. ~\cite{mari_discontinuous_2015} with different strength of repulsion amplitude $F_0$ for $\phi=0.5$. Solid line: model prediction for $\phi=0.5$ with $\sigma_{\rm y}=0$ and $K=1.2$.
}
\label{fig-brownnoncoh}
\end{figure*}

\clearpage

\section*{Normal stress difference with attractive force}
Figure~\ref{fig-N2coh} shows variation of the second normal stress difference $-N_2/\sigma_0$ against shear rate $\dot{\gamma}/\dot{\gamma}_0$. Similar to the shear stress, normal stresses also develop a yield behavior at low shear rates and the behavior is independent of strength of cohesion at high shear rate or shear stress.
We find that the first normal stress difference is smaller compared to the second normal stress difference and is dominated by fluctuations; this data is not presented.
\begin{figure*}[!ht]
\centering
 \includegraphics[width=.45\textwidth]{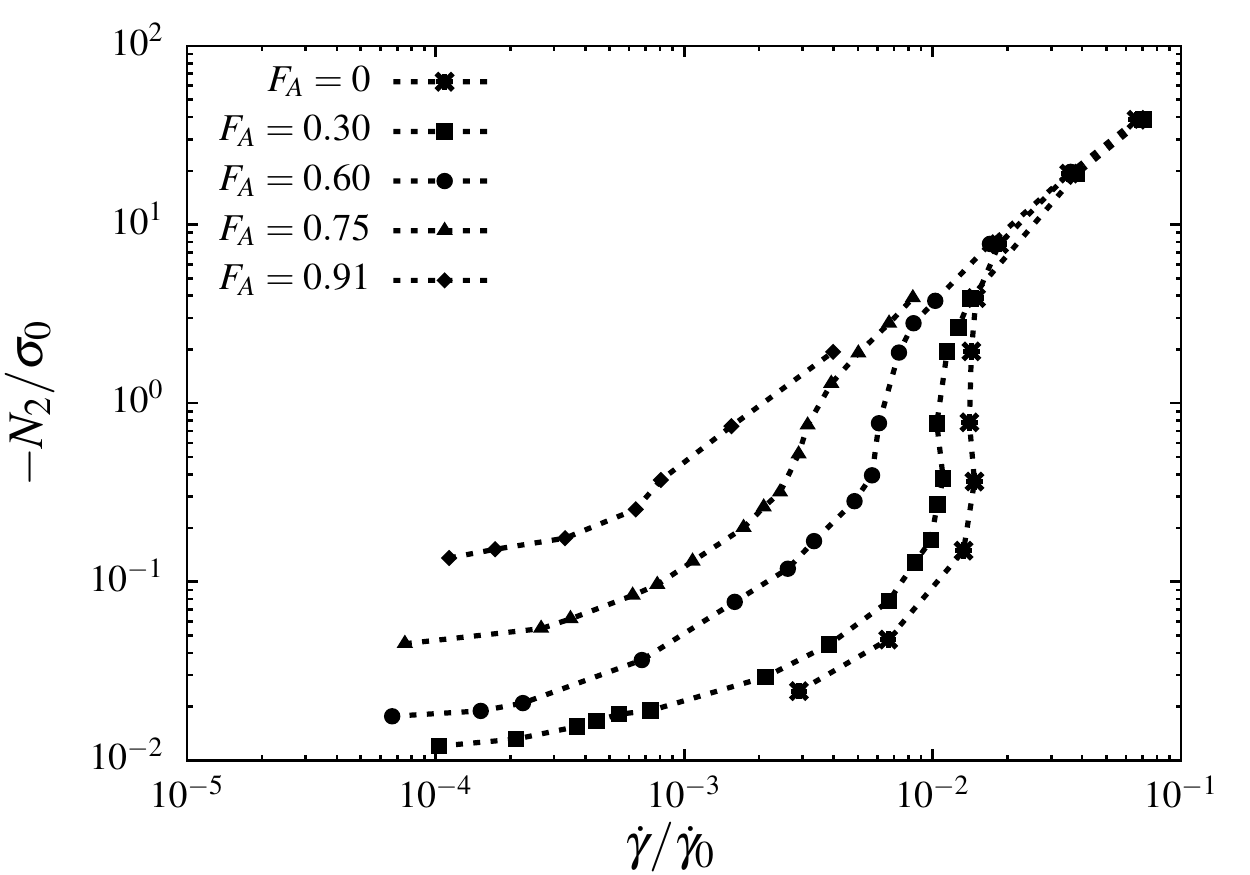}
\caption{
Second normal stress difference $-N_2/\sigma_0$ plotted as a function of shear rate $\dot{\gamma}/\dot{\gamma}_0$ for volume fraction $\phi=0.56$ and several values of attractive strength $F_A$ at $\mu=1$.
}
\label{fig-N2coh}
\end{figure*}

\clearpage

\section*{Flow curves for $F_{\rm A} = 0.3$}
Figure~\ref{fig-eta_FA} shows relative viscosity $\eta_{\rm r}$ plotted as a function of stress $\hat{\sigma}$ for different volume fractions. 
At $\phi=0.45<\phi_{\rm y}^0$ the system flows at vanishing shear stress i.e., no yield stress is observed.
For the range of volume fractions $\phi_{\rm y}^0<\phi<\phi_{\rm J}^\mu$ the suspensions shows a yield stress and flows for $\hat{\sigma}>\hat{\sigma}_{\rm y}$.
For $\phi>\phi_{\rm J}^\mu$ the system shows an unyielded solid state for  $\hat{\sigma}<\hat{\sigma}_{\rm y}$, while it flows for the range of stress 
$\hat{\sigma}_{\rm y}<\hat{\sigma}<\hat{\sigma}_{\rm sj}$.
Above volume fraction $\phi_{\rm {max}}=0.635$ the system cannot flow for any stress.
This information is presented in the form of a state diagram in Fig.~\ref{fig:phase_diagram_fa}.
\begin{figure*}[!ht]
\centering
 \includegraphics[width=.45\textwidth]{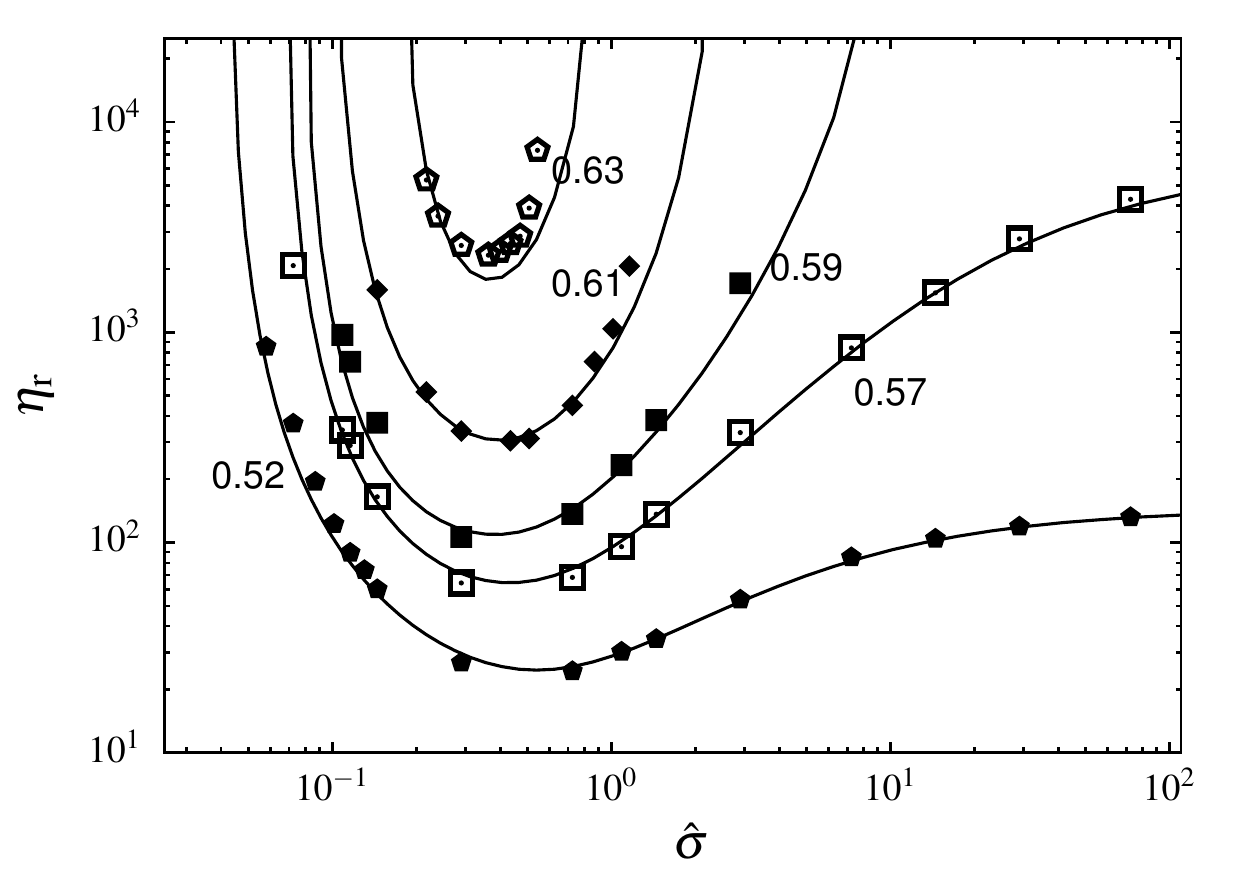}
\caption{
Steady state flow curves for $F_A=0.3$ plotted as a function of scaled applied shear stress $\hat{\sigma}$. The symbols are simulations with different volume fractions $\phi$ and the solid lines are predictions from~\eqref{eq:eta_total}.
}
\label{fig-eta_FA}
\end{figure*}


\end{appendix}
\clearpage

\end{widetext}
\end{document}